\documentclass[prl,superscriptaddress,amsfonts,twocolumn,showpacs,floatfix]{revtex4-1}
\usepackage{amsmath}
\usepackage{amsfonts}
\usepackage{amssymb}
\usepackage{bm}
\usepackage{txfonts}
\usepackage{tabularx}
\usepackage[dvips]{graphicx,color}
\usepackage{dcolumn}

\begin{document}
\title{
Quadrupole Order in the Frustrated Pyrochlore Tb$_{2+x}$Ti$_{2-x}$O$_{7+y}$
}

\author{H. Takatsu}
\affiliation{Department of Physics, Tokyo Metropolitan University, Hachioji-shi, Tokyo 192-0397, Japan}
\affiliation{Department of Energy and Hydrocarbon Chemistry, Graduate School of Engineering, Kyoto University, Kyoto 615-8510, Japan}

\author{S. Onoda}
\affiliation{RIKEN Center for Emergent Matter Science (CEMS), Wako, Saitama 351-0198, Japan}
\affiliation{Condensed Matter Theory Laboratory, RIKEN, Wako, Saitama 351-0198, Japan}

\author{S. Kittaka}
\affiliation{Institute for Solid State Physics, University of Tokyo, Kashiwa 277-8581, Japan}

\author{A. Kasahara}
\affiliation{Institute for Solid State Physics, University of Tokyo, Kashiwa 277-8581, Japan}

\author{Y. Kono}
\affiliation{Institute for Solid State Physics, University of Tokyo, Kashiwa 277-8581, Japan}

\author{T. Sakakibara}
\affiliation{Institute for Solid State Physics, University of Tokyo, Kashiwa 277-8581, Japan}

\author{Y. Kato}
\affiliation{RIKEN Center for Emergent Matter Science (CEMS), Wako, Saitama 351-0198, Japan}
\affiliation{Department of Applied Physics, University of Tokyo, Bunkyo, Tokyo 113-8656, Japan}

\author{B. F\aa k}
\affiliation{Institute Laue Langevin, BP 156, F-38042 Grenoble, France}

\author{J. Ollivier}
\affiliation{Institute Laue Langevin, BP 156, F-38042 Grenoble, France}

\author{J. W. Lynn}
\affiliation{NCNR, National Institute of Standards and Technology, Gaithersburg, MD 20899-6102, U.S.A}

\author{T. Taniguchi}
\affiliation{Department of Physics, Tokyo Metropolitan University, Hachioji-shi, Tokyo 192-0397, Japan}

\author{M. Wakita}
\affiliation{Department of Physics, Tokyo Metropolitan University, Hachioji-shi, Tokyo 192-0397, Japan}

\author{H. Kadowaki}
\affiliation{Department of Physics, Tokyo Metropolitan University, Hachioji-shi, Tokyo 192-0397, Japan}
\date{\today}

\begin{abstract}
A hidden order that emerges in the frustrated pyrochlore 
Tb$_{2+x}$Ti$_{2-x}$O$_{7+y}$ with $T_{\text{c}}=0.53$ K is 
studied using specific heat, magnetization, 
and neutron scattering experiments on a high-quality single crystal. 
Semi-quantitative analyses based on a pseudospin-1/2 Hamiltonian 
for ionic non-Kramers magnetic doublets demonstrate that 
it is an ordered state of electric quadrupole moments. 
The elusive spin liquid state of the nominal 
Tb$_2$Ti$_2$O$_7$ is most likely a U(1) quantum spin-liquid state.
\end{abstract}

\maketitle

\begin{figure}[t]
\begin{center}
 \includegraphics[width=0.475\textwidth]{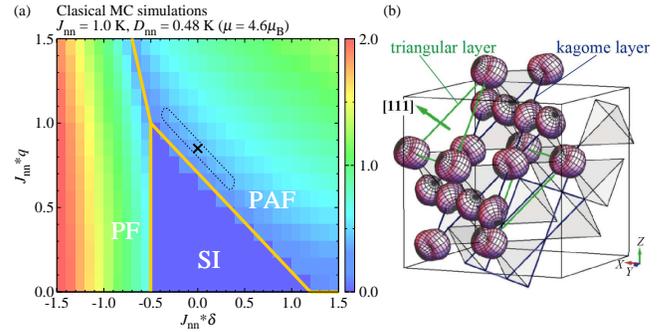}
\caption{
(Color online)
(a) Phase diagram of the effective Hamiltonian Eq.~(\ref{eq.1}) 
determined from CMC simulations. The shading (color) represents $T_{\text{c}}$. 
{\color[rgb]{0,0,0} Two quadrupole LRO phases, 
the planar antiferropseudospin (PAF) and planar ferropseudospin (PF) phases, 
exist in the vicinity of the SI phase~\cite{S.Onoda2011PRB}.}
Classical SI is replaced by a U(1) QSL in quantum theory~\cite{LeePRB2012}.
The region enclosed by the dotted line represents an acceptable parameter region for 
the experimental data on Tb$_{2.005}$Ti$_{1.995}$O$_{7+y}$.
The cross mark indicates the typical values $(\delta, q) = (0,0.85)$.
(b) Schematic view of the deformation of the $f$-electron charge density 
due to the PAF order on the pyrochlore lattice.
}
\label{fig.1}
\end{center}
\end{figure}

Geometrically frustrated magnets have been actively investigated 
in condensed matter physics~\cite{C.Lacroix}.
In particular, 
spin ice (SI), e.g. $R$$_2$Ti$_2$O$_7$ ($R=$ Dy or Ho) 
\cite{HarrisPRL1997,BramwellScience2001}, 
provides prototypical frustrated Ising magnets with the pyrochlore lattice structure \cite{GardnerRMP2010}, 
consisting of a three-dimensional network of corner-sharing tetrahedra [Fig.~\ref{fig.1}(b)].
It displays 
fascinating features such as a finite zero-point entropy \cite{RamirezNature1999} 
and thermally excited emergent magnetic or SI monopoles \cite{RyzhkinJETP2005,CastelnovoNature2008}. 
An intriguing theoretical proposal for a U(1) quantum spin liquid (QSL) state \cite{HermelePRB2004} 
has been made for variants of SI endowed with quantum spin
fluctuations~\cite{LeePRB2012,SavaryPRL2012,S.Onoda2011PRB,HuangPRL2014,GingrasRPP2014,MolavianPRL2007}. 
The U(1) QSL state~\cite{HermelePRB2004,SavaryPRL2012,LeePRB2012} 
is characterized by an emergent U(1) gauge field producing gapless fictitious photons
and by gapped bosonic spinon excitations carrying the SI magnetic monopole 
charge~\cite{HermelePRB2004,BentonPRB2012,LeePRB2012,GingrasRPP2014}. 
By increasing the transverse interaction, the system can undergo 
a phase transition from the U(1) QSL to a long range ordered (LRO) state of 
transverse spins or pseudospins representing electric-quadrupole moments for non-Kramers ions~\cite{LeePRB2012,SavaryPRL2012,S.Onoda2011PRB}.
This state can be described as a Higgs phase~\cite{NambuPR1960,P.W.HiggsPRL1964,P.W.AndersonPR1963,FradkinPRD1979,RokhsarPRL1988}. 

In a quest to QSL states in frustrated magnetic systems from both theoretical~\cite{Anderson1973, LeeScinence2008, BalentsNature2010}
and experimental~\cite{ShimizuPRL2003,HanNature2012} viewpoints, 
an Ising-like pyrochlore Tb$_{2}$Ti$_{2}$O$_{7}$ (TTO) is a potential candidate for a U(1) QSL: 
it has been reported to remain in a fluctuating spin state down to 50~mK without
magnetic LRO~\cite{GardnerPRL1999,GardnerPRB2003}.
However, the origin of this spin liquid state of TTO has been 
elusive for more than a decade despite many investigations (see Refs.~\cite{GardnerRMP2010, GingrasRPP2014,PetitEPJWC2015} 
and references therein, and recent Refs.~\cite{HirschbergerScience2015,PrincepPRB2015,GuittenyPRB2015}), 
and is still under hot debate~\cite{GingrasRPP2014,PetitEPJWC2015}. 
To solve this challenging problem of TTO, 
we start this investigation by postulating that 
the theoretically-proposed interaction between 
electric quadrupole moments of non-Kramers ions including Tb$^{3+}$ 
[the fourth term of Eq.~(\ref{eq.1})] \cite{S.Onoda2011PRB} 
is at work for giving the quantum fluctuations to TTO. 
This postulation is a natural consequence of 
the previous unsuccessful trial-and-errors of explaining TTO by taking into account only
the interactions between magnetic dipole moments [the first three terms of Eq.~(\ref{eq.1})] 
and the perturbation through first excited crystal-field (CF) states \cite{KaoPRB2003,MolavianPRL2007}, 
and by taking another assumption of Jahn-Teller (JT) distortion \cite{BonvillePRB2011,PetitEPJWC2015}. 
Under the present postulation, two ground states of off-stoichiometric 
Tb$_{2+x}$Ti$_{2-x}$O$_{7+y}$ samples~\cite{TaniguchiPRB2013}
will possibly be accounted for by the U(1) QSL ($x < x_{\text{c}}$) and 
electric quadrupolar ($x > x_{\text{c}}$) states of Ref.~\cite{LeePRB2012}.

In this Letter, 
we investigate the hidden order of 
Tb$_{2+x}$Ti$_{2-x}$O$_{7+y}$ ($x = 0.005 > x_{\text{c}}$), 
because the electric quadrupolar order is more tractable than the U(1) QSL by 
using semi-classical theoretical analyses. 
Specific heat, magnetization, and neutron scattering experiments 
were performed, and these experimental data were analyzed 
using quantum and classical Monte Carlo (QMC, CMC) simulations, 
and a mean-field random-phase approximation (MF-RPA). 
The results demonstrate that the hidden order is an electric quadrupolar order [Fig.~\ref{fig.1}(b)]
and that the parameters of the model Hamiltonian are located close to a phase boundary 
between the electric quadrupole and U(1) QSL states [Fig.~\ref{fig.1}(a)],
which suggests that the elusive spin-liquid state of TTO is the U(1) QSL. 
We emphasize that a high-quality single-crystalline sample 
with a well-controlled $x$ value~\cite{M.WakitaJPCS2016} 
enables us to accomplish this work.

An effective pseudospin-1/2 Hamiltonian \cite{S.Onoda2011PRB} relevant 
for the non-Kramers magnetic doublets of TTO is 
\begin{alignat}{3}
\mathcal{H} = &J_{\rm nn}\sum_{\langle \bm{r}, \bm{r}' \rangle}\sigma_{\bm{r}}^{z}\sigma_{\bm{r}'}^{z}  - \mu_{\text{eff}} \bm{H} \cdot \sum_{\bm{r}} \bm{e}_{\bm{r}}^{z} \sigma_{\bm{r}}^{z} \notag \\
         +& Dr_{\rm nn}^3\sum \Biggl[ \frac{{\bm{e}}_{\bm{r}}^z\cdot {\bm{e}}_{\bm{r}'}^z}{|\bm{r}-\bm{r}'|^3} - \frac{3[ {\bm{e}}_{\bm{r}}^z \cdot (\bm{r}-\bm{r}') ][ {\bm{e}}_{\bm{r}'}^z \cdot (\bm{r}-\bm{r}') ]}{|\bm{r}-\bm{r}'|^5} \Biggr]\sigma_{\bm{r}}^{z}\sigma_{\bm{r}'}^{z} \notag \\
         +&J_{\rm nn}\sum_{\langle \bm{r}, \bm{r}' \rangle} \Bigl[ 2\delta \bigl(\sigma_{\bm{r}}^{+} \sigma_{\bm{r'}}^{-} + \sigma_{\bm{r}}^{-}\sigma_{\bm{r'}}^{+} \bigr) + 2q \bigl( e^{i2\phi_{\bm{r},\bm{r'}}} \sigma_{\bm{r}}^{+}\sigma_{\bm{r'}}^{+} + \rm{H.c.} \bigr)\Bigr]. 
\label{eq.1}
\end{alignat}
Here, we consider only the CF ground state doublet~\cite{MirebeauPRB2007,PrincepPRB2015}, 
and neglect the first excited doublet at $E \simeq 18$~K, 
since we are mainly interested in the low-$T$ properties below 2~K. 
In Eq.~(\ref{eq.1}), $\bm{\sigma}_{\bm{r}}$ are the Pauli matrices (pseudospin) at a site $\bm{r}$, 
$\sigma_{\bm{r}}^{\pm} \equiv (\sigma_{\bm{r}}^{x}\pm i \sigma_{\bm{r}}^{y})/2$, 
and $\phi_{\bm{r},\bm{r'}} = \pm \tfrac{2\pi}{3}, 0$ \cite{S.Onoda2011PRB,KadowakiSPIN2015}. 
The magnetic dipole moment $\mu_{\text{eff}} \sigma_{\bm{r}}^{z}$ is parallel to 
the local $\langle 111 \rangle$ axis $\bm{e}_{\bm{r}}^z$ \cite{KadowakiSPIN2015}. 
The first three terms of Eq.~(\ref{eq.1}) represent the nearest-neighbor (NN) exchange interaction, the Zeeman energy under a magnetic field $\bm{H}$, 
and the dipolar interaction, respectively. 
They constitute the classical dipolar SI Hamiltonian $\mathcal{H}_{\text{m}}$ \cite{HertogPRL2000}.
It can be approximated \cite{HertogPRL2000,IsakovPRL2005} by the NN classical SI model 
$\mathcal{H}_{\text{m,eff}}=J_{\text{nn,eff}} \sum_{\langle \bm{r} , \bm{r}^{\prime} \rangle} \sigma_{\bm{r}}^{z} \sigma_{\bm{r}^{\prime}}^{z} - \mu_{\text{eff}} \bm{H} \cdot \sum_{\bm{r}} \bm{e}_{\bm{r}}^{z} \sigma_{\bm{r}}^{z}$, 
where $J_{\text{nn,eff}}= J_{\text{nn}} + D_{\text{nn}}$ ($D_{\text{nn}} = \tfrac{5}{3} D$). 
The last term  of Eq.~(\ref{eq.1}) represents the quadrupole interaction $\mathcal{H}_{\text{q}}$. 
We note that the transverse components $(\sigma_{\bm{r}}^{x}, \sigma_{\bm{r}}^{y})$ of the pseudospin 
represent electric quadrupole (and 16-, 64-pole) moments~\cite{S.Onoda2011PRB,KadowakiSPIN2015}. 
{\color[rgb]{0,0,0} Note that QMC simulations of the model Eq.~(\ref{eq.1}) suffer from a negative sign problem.
On the other hand, thermodynamic properties away from the QSL state, including phase transitions to the LRO phases,
can be captured by CMC simulations semi-quantitatively. Therefore, in most of the cases, we employ CMC simulations.}

CMC simulations were performed up to 1024 pseudospins, in which 
the pseudospin $\bm{\sigma}_{\bm{r}}$ is treated as a classical unit vector~\cite{Kadowaki_MC}.
The resulting zero-field phase diagram is shown in Fig.~\ref{fig.1}(a) for
the case of $D_{\mathrm{nn}}=0.48$~K, namely, $\mu=4.6\mu_B$~\cite{Supplemental_TTO}, and $J_{\mathrm{nn}}=1.0$~K
(this value of $D_{\mathrm{nn}}$ will be used throughout the paper
and the choice of $J_{\mathrm{nn}}$ will be explained further below).
A quantum mechanical treatment using gauge mean-field (MF) theory shows 
that the classical SI phase region in Fig.~\ref{fig.1}(a) is mostly replaced by 
a U(1) QSL phase except at $\delta = q =0$~\cite{LeePRB2012}.
The phase diagram has two quadrupole LRO phases
originating from different ordering patterns of $(\sigma_{\bm{r}}^{x}, \sigma_{\bm{r}}^{y})$:
PAF (planar antiferropseudospin) and PF (planar ferropseudospin) states 
denoted in the classical MF phase-diagram (Fig.~7 in Ref.~\cite{S.Onoda2011PRB}). 
In particular, a deformation of the $f$-electron charge density \cite{KusunoseJPSJ2008} for 
the PAF phase is illustrated in Fig.~\ref{fig.1}(b)~\cite{LeePRB2012,Kadowaki_MC}.
In the following, we will show that most of the experimental data on 
the high-quality polycrystalline and single-crystalline samples of TTO with $x=0.005$ can be explained by
choosing $J_{\text{nn}}=1$~K, $\delta=0$, and $q=0.85$ within semi-quantitative analyses. 

%

Polycrystalline and single-crystalline Tb$_{2+x}$Ti$_{2-x}$O$_{7+y}$ 
samples were prepared 
by a standard solid-state reaction~\cite{TaniguchiPRB2013} and 
by a floating zone method~\cite{M.WakitaJPCS2016}. 
Specific heat was measured by a quasi-adiabatic method down to 0.1~K 
using a plate-shape crystal with a size of $0.7\times0.9\times0.1$~mm$^3$
whose shortest dimension is along a [110] axis.
Magnetization was measured by a capacitive Faraday magnetometer 
using the same sample. 
Neutron scattering experiments were performed on 
NIST-BT7~\cite{LynnJNIST2012} using a crystal sample cut from a neighboring part of the above sample
and on ILL-IN5 \cite{OllivierJPSJ2011} using the powder sample with $x=0.005$~\cite{TaniguchiPRB2013}.

\begin{figure}[t]
\begin{center}
 \includegraphics[width=0.44\textwidth]{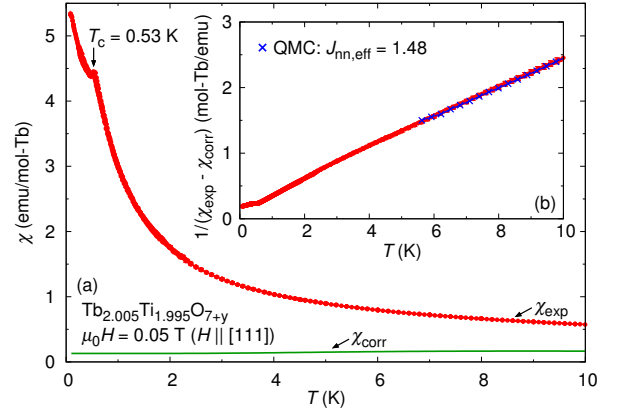}
\caption{
(Color online)
(a) Temperature dependence of magnetic susceptibility for an applied field along [111]. 
The solid line is $\chi_{\text{corr}}$ (see text for details).
(b) Comparison between $[\chi_{\text{exp}}-\chi_{\text{corr}}]^{-1} = \chi_{\rm Gnd}^{-1}$ and QMC calculation. 
Note, 1~emu = 10$^{-3}$~A m$^{2}$.
}
\label{fig.2}
\end{center}
\end{figure}
\begin{figure*}[t]
\begin{center}
 \includegraphics[width=0.90\textwidth]{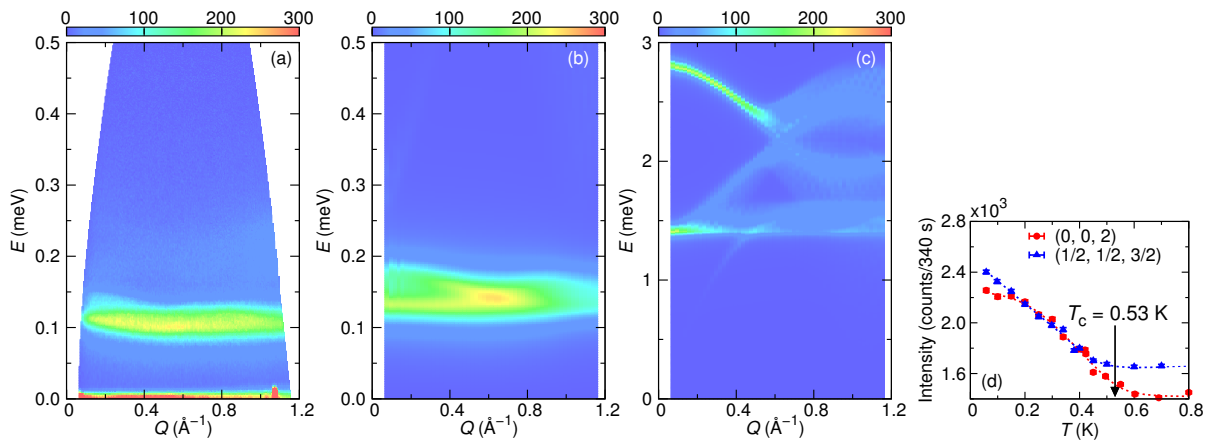}
\caption{
(Color online) 
(a) Neutron inelastic magnetic spectra of the polycrystalline 
Tb$_{2.005}$Ti$_{1.995}$O$_{7+y}$ sample taken at $T = 0.1$~K. 
(b) Calculated $S(Q, E)$ for the PAF phase with $J_{\text{nn,eff}} > 0$
using 
$J_{\rm nn,eff}=1.48$~K ($J_{\rm nn} = 1.00$~K) 
and $(\delta, q) = (0,0.85)$. 
$J_{\rm nn,eff}$ is determined by the analysis of $\chi_{\rm Gnd}$.
(c) Calculated $S(Q, E)$ for the PAF phase with $J_{\text{nn,eff}} < 0$
using
$J_{\rm nn,eff}=-1.77$~K ($J_{\rm nn} =-2.25$~K)
and $(\delta, q) = (-0.5,-1.0)$.
$J_{\rm nn,eff}$ is determined by the analysis of $\chi_{\rm Gnd}$.
(d) 
$T$-dependence of intensities of the single-crystal neutron Bragg scattering 
at $(0 0 2)$ and ($\frac{1}{2} \frac{1}{2} \frac{3}{2}$). 
}
\label{fig.3}
\end{center}
\end{figure*}

We first determine the magnitude of $J_{\text{nn}}$ from 
the magnetic susceptibility $\chi_{\rm exp}$.
The $T$ dependence of $\chi_{\rm exp}$ measured along the [111] direction on 
the single crystal shows an anomaly at $T_{\mathrm{c}}=0.53$~K [Fig.~\ref{fig.2}(a)].
While $\chi_{\mathrm{exp}}$ is dominated by the contribution from 
the CF ground-state doublets, a small but non-negligible correction
$\chi_{\text{corr}}$ may arise from higher-energy CF states.
Thus, we calculated $\chi_{\text{corr}}$ 
by taking the CF parameters of Ref.~\cite{MirebeauPRB2007} and using a single-site approximation, 
i.e., using Eq.(2.1.18) of Ref.~\cite{Jensen91} 
where the contributions from the CF ground state doublet are excluded. 
The $\chi_{\text{corr}}$ is also shown in Fig.~\ref{fig.2}(a).
Now $\chi_{\rm Gnd} (= \chi_{\rm exp} - \chi_{\text{corr}})$ can be directly compared with 
a theoretical calculation based on the pseudospin-1/2 model Eq.~(1). 
We have performed extensive QMC simulations \cite{Y.KatoPRL2015,KatoPRE2009} 
of the nearest-neighbor effective Hamiltonian $\mathcal{H}_{\text{m,eff}} + \mathcal{H}_{\text{q}}$ 
on finite-size clusters up to 1024 pseudospins with typical Monte-Carlo steps of 200000. 
The experimental data ($\chi_{\rm Gnd}$) are well reproduced by the QMC calculations
in a wide range of $\delta$ and $q$, 
if we take $|J_{\text{nn,eff}}| = 1.3 - 1.9$~K. 
Note that because of the negative sign problem of the QMC simulation, the analyses 
have been limited to a relatively high temperature range, $5<T<15$~K.
In Fig.~\ref{fig.2}(b), we show a representative comparison between $\chi_{\rm Gnd}$
and the QMC results obtained for
$(\delta,q) = (0, 0.85)$ as determined below. 
This comparison yields $J_{\text{nn,eff}} = 1.48(1)$,
leading to $J_{\rm nn} = 1.0(1)$~K. 
This value of $J_{\text{nn,eff}}$ 
is of the same order as the previous estimation~\cite{GingrasPRB2000}.

Next, we confirm the positive sign of $J_{\text{nn,eff}}$ and extract the parameter values of 
$(\delta,q)$ from the comparison of the inelastic neutron scattering data.
The previous inelastic magnetic neutron-scattering spectra measured on the $x=0.005$ powder sample at 
$T = 0.1$~K ($\ll T_{\text{c}}$), have shown a nearly flat broad peak at 0.1~meV 
in the $(Q,E)$ space \cite{TaniguchiPRB2013}, as shown in Fig.~3(a).
{\color[rgb]{0,0,0}The peak is broader than the instrumental resolution, which suggests dispersive excitations.}
This experimental behavior can be described in terms of
pseudospin waves in the PAF and PF phases,
as discussed in Ref.~\cite{KadowakiSPIN2015}.
The powder-averaged dynamical magnetic structure factor
$S(|\bm{Q}|, E)$ is calculated within the MF-RPA~\cite{KadowakiSPIN2015},
{\color[rgb]{0,0,0} which can correctly describe the spectrum in an ordered state
within the linear spin-wave approximation.}
Extensive calculations in a wide range of the parameters $(\delta, q)$
show that reasonable agreements are obtained in 
the PAF phase with $(\delta,q) = (0.0\pm0.4, 0.8\pm0.3)$ [Fig.~\ref{fig.3}(b)]
and in the PF phase with $(\delta,q) = (-0.54\pm0.02, q<1.1)$, 
when we fix $J_{\rm nn} = 1.0(1)$~K [Fig.~\ref{fig.1}(a)].
We note that only the cases of $J_{\text{nn,eff}} > 0$ can 
reasonably reproduce the observed features for the case of the PAF phase [Fig.~\ref{fig.3}(b)].
The case of $J_{\mathrm{nn,eff}}<0$ gives highly dispersive spectra that 
are not compatible with the experimental results [Fig.~\ref{fig.3}(c)].

\begin{figure*}[t]
\begin{center}
 \includegraphics[width=0.90\textwidth]{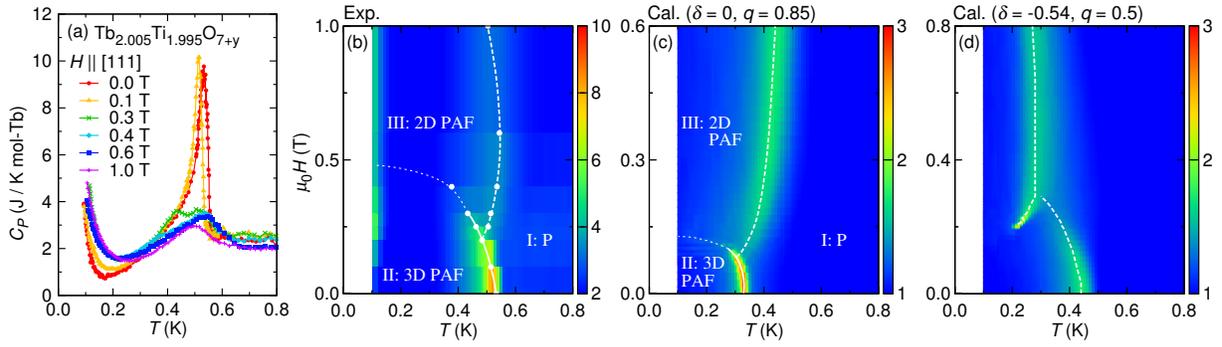}
\caption{
(Color online)
(a) Temperature dependence of the observed specific heat $C_P(T, H)$ for $H\parallel[111]$.
(b) Temperature-field map of $C_P(T, H)$ for $H\parallel[111]$.
Filled circles in the map are peak positions of $C_P(T, H)$.
(c,d) CMC results of specific heat $C(T, H)$ 
for (c) $(\delta, q) = (0, 0.85)$ 
with $J_{\rm nn} = 1.00$~K and 
for (d) $(\delta, q) = (-0.54, 0.5)$
with $J_{\rm nn} = 0.92$~K. 
The values of $J_{\rm nn}$ have been determined from the comparison of $\chi_{\rm Gnd}$ with the QMC results for each case.
Solid, dashed, and dotted lines in (b), (c), and (d) are guides to the eyes.
{\color[rgb]{0,0,0} Labels in maps (b) and (c) indicate assigned states from the analysis with Eq.(\ref{eq.1}) and CMC simulations~\cite{Kadowaki_MC}; 
i.e., (I) a paramagnetic paraquadrupole state, (II) the 3D PAF state, and (III) the 2D PAF state.}
}
\label{fig.4}
\end{center}
\end{figure*}

Finally, we show, using CMC simulations of Eq.~(1), 
that the parameter set showing the PAF
explain reasonably well the observed specific heat $C_P(T,H)$
under weak [111] field
while the other parameter set showing the PF does not.
Figure~\ref{fig.4}(a) shows the $T$ dependence of $C_P(T,H)$ under 
[111] field up to 1.0~T. The sharp peak at $T_{\mathrm{c}}$ survives only up to $\mu_{0}H=0.1$~T, 
turning into broad double peaks at 0.3~T. A full map of $C_P(T,H)$ is shown in Fig.~\ref{fig.4}(b). 
For comparison, maps of the calculated specific heat $C(T,H)$ by CMC simulations 
are presented in Figs.~\ref{fig.4}(c) and (d) for the same parameter choices as determined above, namely, 
$(\delta,q)=(0,0.85)$ and $(-0.54,0.5)$, respectively.
Clearly, the PAF case shows a better qualitative agreement with the experiment, 
although with some discrepancy in the magnetic field and temperature scales.
{\color[rgb]{0,0,0}
We note that CMC simulations with these PAF parameters also reproduce the experimental 
results of $C(T,H)$ under the $[100]$ field~\cite{H.TakatsuJPCS2016}.}

\begin{figure}[t]
\begin{center}
 \includegraphics[width=0.44\textwidth]{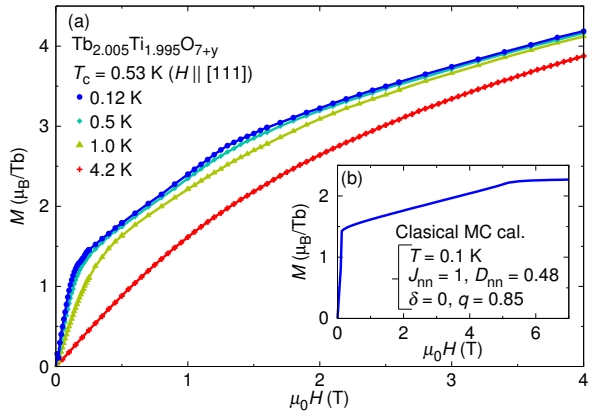}
\caption{
(Color online)
Field dependence of the magnetization $M$ under the [111] magnetic field. 
(a) Experimental data on Tb$_{2.005}$Ti$_{1.995}$O$_{7+y}$
above and below $T_{\rm c} = 0.53$~K.
(b) 
CMC results obtained with the same parameters as 
for Figs.~\ref{fig.3}(b) and \ref{fig.4}(c).
}
\label{fig.5}
\end{center}
\end{figure}
%
{\color[rgb]{0,0,0}One may not clearly see the change of states under the [111] field
below 0.4~K in $C_P(T, H)$ [Figs.~\ref{fig.4}(a) and (b)].
However, certain changes are observed in the magnetization $M$.
Figure~\ref{fig.5}(a) shows $M$--$H$ curves at several temperatures under the [111] field.
Two clear step-like kinks are observed at $\mu_0H_{1} \simeq 0.14$~T and $\mu_0H_{2} \simeq 1.3$~T below the zero-field $T_{\text{c}}$.
CMC simulations with the same parameters as used above demonstrate 
the first kink at $\mu_0H_{1}$ [Fig.\ref{fig.5}(b)],
indicating that it is a crossover or a phase transition from 
the three dimensional (3D) PAF state~\cite{LeePRB2012} to the two dimensional (2D) PAF state~\cite{S.Onoda2011PRB,Kadowaki_MC}.
This means that 
in intermediate fields ($\mu_0H \simeq 0.5$~T), the system behaves as decoupled 2D kagom\'{e} layers of 
quadrupole moments separated by triangular layers of polarized magnetic moments.
This bears resemblance to the kagom\'{e} ice state of SI materials~\cite{MatsuhiraJPCM2002}.
In contrast, 
the second kink appears at a higher field than that of experiments at $\mu_0H_{2}$.
This result suggests that 
higher order terms neglected in Eq.~(\ref{eq.1}), such as terms due to CF excited states~\cite{Supplemental_TTO},
are required for further explanation of the behavior at fields higher than about 1~T.}

All the above comparisons between the experiments and theories show that 
the LRO of TTO is the quadrupole order characterizing the PAF phase.
{\color[rgb]{0,0,0}Although this LRO cannot be detected directly through neutron Bragg scattering, 
we observed some indication of it. 
In fact, weak magnetic reflections are observed at the forbidden $(002)$ position 
and at the superlattice $(\frac{1}{2} \frac{1}{2} \frac{3}{2})$ position [Fig.~\ref{fig.3}(d)]. 
Polarized neutron scattering experiments at BT7 confirm that both of them are magnetic. 
The long-range ordered magnetic moments of these reflections are roughly 
$\sim 0.1$~$\mu_{\text{B}}$, which is too small to be the primary order parameter. 
We speculate that the $(002)$ reflection appears simultaneously 
with the PAF order, whose order parameter is characterized by 
the wave vector $\bm{k} = 0$~\cite{LeePRB2012,Kadowaki_MC}, 
and is induced by higher order terms neglected in Eq.~(\ref{eq.1}). 
On the other hand the $(\frac{1}{2} \frac{1}{2} \frac{3}{2})$ reflection, 
observed also in a powder sample~\cite{TaniguchiPRB2013}, 
suggests a different origin because of the different $T$-dependence 
for the $(002)$ reflection [Fig.~\ref{fig.3}(d)].}

{\color[rgb]{0,0,0} Electric quadrupolar orders are related to the deformation of $f$-electron charge density.
These naturally couple to displacements of ligand ions~\cite{Jensen91} and may induce cooperative JT effects
and JT structural distortions~\cite{BonvillePRB2011,GehringRPP1975}.
Since the quadrupole-coupling terms of Eq.~(\ref{eq.1}) are derived as the electronic coupling
but are symmetry allowed terms under the space group of the pyrochlore lattice~\cite{S.Onoda2011PRB},
these may contain a phonon-coupling contribution~\cite{SazonovPRB2013}.
Thus, a direct detection of the quadrupole order (using resonant X-ray scattering) 
and/or of a small associated JT lattice distortion are difficult yet interesting topics for future investigations.}

In summary, the hidden order of Tb$_{2+x}$Ti$_{2-x}$O$_{7+y}$ with $x=0.005>x_{\text{c}}$ has been studied using
thermodynamic and neutron scattering measurements on single-crystalline and polycrystalline samples 
under the control of the off-stoichiometry of $x$.
We take account of magnetic-dipole and electric-quadrupole moments of the CF ground-state doublet of 
the non-Kramers Tb$^{3+}$ ion as well as the theoretically proposed quadrupole interaction~\cite{S.Onoda2011PRB}.
Semi-quantitative analyses of the experimental data based on a simple pseudospin-1/2 Hamiltonian demonstrate that 
the hidden order is an order of the electric quadrupole moments [Fig.~\ref{fig.1}(b)].
The estimated model parameters are located close to the phase boundary between the quadrupolar and U(1) QSL states.
This result implies that the putative SL state of TTO studied for more than a decade is the U(1) QSL.
{\color[rgb]{0,0,0}Investigations in the context of a Higgs transition~\cite{LeePRB2012,SavaryPRL2012,ChangNatureComm2012}
and on the relation of it with the neighboring U(1) QSL phase using Tb$_{2+x}$Ti$_{2-x}$O$_{7+y}$ single crystals~\cite{M.WakitaJPCS2016}
are fascinating future topics.}

\begin{acknowledgments}
We acknowledge K. Matsuhira, M. J. P. Gingras, B. D. Gaulin, K. Matsubayashi, and 
R. Higashinaka for fruitful discussion.
This work was supported by JSPS KAKENHI Grant Numbers 25400345, 26400336, and 26800199.
The specific heat and magnetization measurements were
performed using the facilities of ISSP, Univ. of Tokyo.
The neutron scattering performed using ILL-IN5 (France) 
was transferred from JRR3-HER (proposal 11567) 
with the approval of ISSP, Univ. of Tokyo, and JAEA, Tokai, Japan.
Numerical calculations were conducted on RICC and HOKUSAI-GW.
\end{acknowledgments}

%
\bibliography{20160417_TTO_PRL_paper_B.bbl}

\pagebreak
\widetext
\begin{center}
\textbf{\large Supplemental Material: Quadrupole Order in the Frustrated Pyrochlore Tb$_{2+x}$Ti$_{2-x}$O$_{7+y}$}
\end{center}
\setcounter{equation}{0}
\setcounter{figure}{0}
\setcounter{table}{0}
\setcounter{page}{1}
\makeatletter
\renewcommand{\theequation}{S\arabic{equation}}
\renewcommand{\thefigure}{S\arabic{figure}}
\renewcommand{\bibnumfmt}[1]{[S#1]}
\renewcommand{\citenumfont}[1]{S#1}

\section{Abstract}

{\color[rgb]{0,0,0}In this supplemental material,
we discuss effects of higher excited crystal field (CF) states of Tb$_2$Ti$_2$O$_7$ (TTO) 
using the analysis of the magnetization $M$.
We also present the result of inelastic neutron scattering (INS) data above $T_{\rm c}$
and discuss a local symmetry reduction which couples to the static Jahn-Teller distortion
but is thought to be a very minor effect for the present system.}


\section{Effects of higher excited CF states: $M$-$H$ curve under [111] field}
{\color[rgb]{0,0,0}When we consider TTO in the wider $T$ and wider $H$ ranges beyond the scope of the present investigations, 
we cannot neglect the contribution from the first excited CF doublet state ($E\simeq18$~K)~\cite{GingrasPRB2000_S}. 
An interesting quantum effect coming from the first excited CF state was pointed out in Ref.~\cite{MolavianPRL2007_S}. 
The NN exchange couplings between the magnetic dipole moments ($\bm{J}_i \cdot\bm{J}_j$) gives rise to a perturbative term 
to the effective $S=1/2$ Hamiltonian in the same form as the $\delta \neq 0$ term in Eq.~(1) of the main text. 
Thus the effective pseudospin $S=1/2$ Hamiltonian that we postulate in this work (Eq.~(1) in the main text) 
may be regarded as a renormalized $S=1/2$ Hamiltonian including effects of the higher excited CF states.

In this section, we try to explain how the higher excited CF states 
affect the present system from an experimental viewpoint. 
To this end the magnetization curve under the [111] magnetic field is a good example. 
A magnetization curve within a single-site approximation can be calculated using the single-site CF Hamiltonian~\cite{Jensen91_S}. 
We thus used the CF parameter set given in Ref.~\cite{MirebeauPRB2007_S} 
and took account of the four crystallographic sites. 

The $M$-$H$ curve at 0.1 K is calculated and plotted in Fig.~\ref{fig.s2}. 
In this figure, we also plotted the experimentally observed $M$-$H$ curve at 0.12 K
(the same data in Fig.~5(a) of the main text) for comparison.
The obvious discrepancy between these two $M$-$H$ curves 
should be accounted for by multi-site effects, i.e., by quantum or 
classical many-body effects of this frustrated system. 
Note that the gradual increase of the calculated 
$M$ in $\mu_0H > 0.5$~T originates mainly from 
the hybridization of the first excited CF states to the CF ground states. 
This is a simple single-site quantum effect. 
A similar gradual increase is also observed in the experimental $M$-$H$ curve.
It is thus considered that a useful $H$ range of the effective $S=1/2$ Hamiltonian 
(Eq.~(1) of the main text) is $\mu_0H<1$~T.}

\begin{figure}[h]
\begin{center}
 \includegraphics[width=0.44\textwidth]{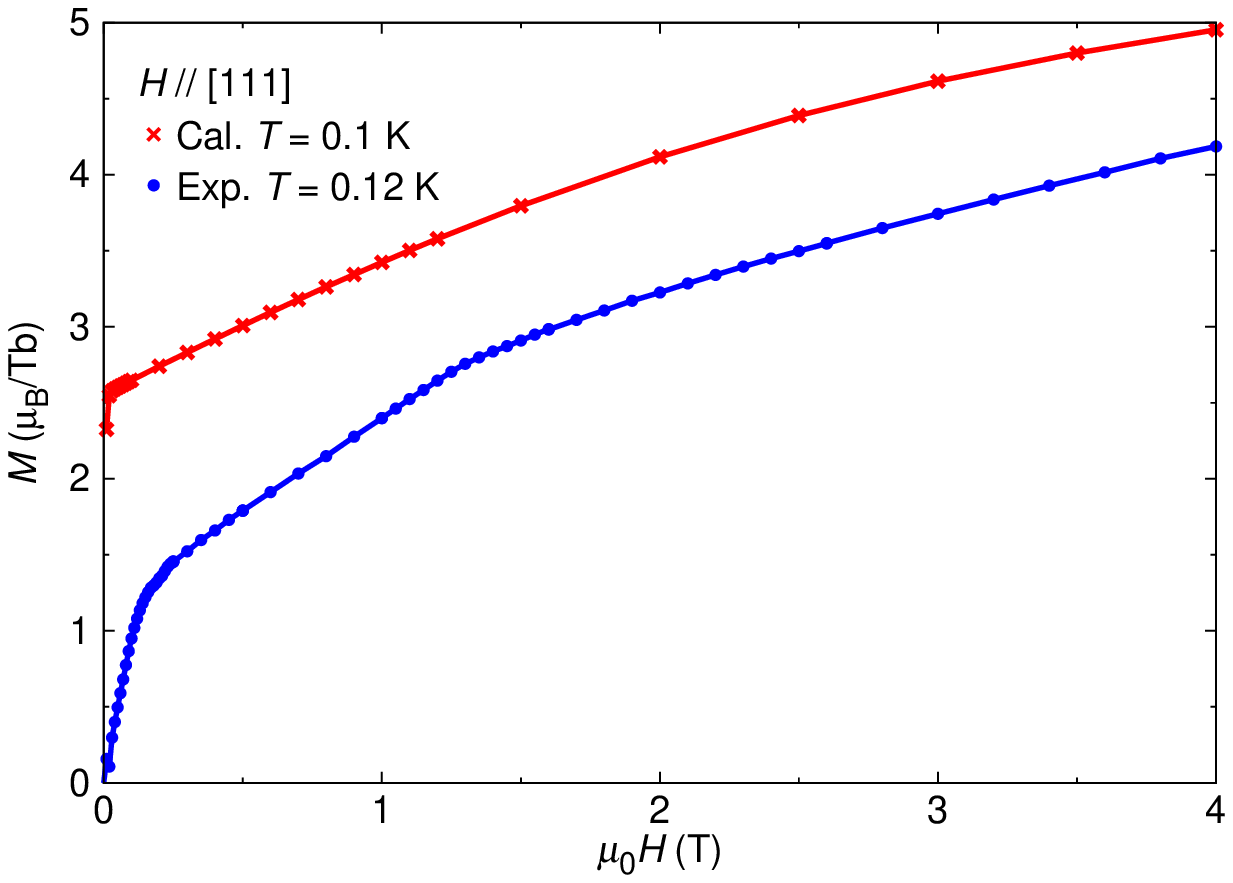}
\caption{
(Color online)
Magnetization curves under the [111] magnetic field at 0.1~K. 
The calculated $M$-$H$ curve within the single-site approximation and 
the experimentally observed $M$-$H$ curve at 0.12~K are shown.
}
\label{fig.s1}
\end{center}
\end{figure}
{\color[rgb]{0,0,0}In Fig.~\ref{fig.s1}, the calculated $M$-$H$ curve exhibits a quick rise to 
$M \simeq 2.6~\mu_{\text{B}}$ in $\mu_0H<0.05$~T. 
This is naturally considered as that the magnetic dipole moment of 
Tb$^{3+}$ is fully polarized along a local [111] axis 
within the CF ground state doublet in low field. 
On the other hand, the experimental $M$-$H$ curve exhibits 
a slower increase around $\mu_0H \simeq 0.1$~T and the kink at $\mu_0H_1\simeq0.14$~T ($M \simeq 1.2$~$\mu_{\text{B}}$/Tb). 
The classical Monte Carlo (CMC) simulation shown in Fig.~5(b) of the main text
approximately reproduces these experimental behaviors. 
We can thus think a classical many-body effect is at work around the kink to certain extent. 
Around the first kink of $\mu_0H_1$ for the CMC simulation, 
the magnetization is about $M = 1.5~\mu_{\text{B}}/{\text{Tb}} = (4.6/3)~\mu_{\text{B}}/{\text{Tb}}$.
This implies that the state bears certain resemblance to the 1/3-plateau state or 
the kagom\'{e} ice state of SI materials~\cite{MatsuhiraJPCM2002_S,TabataPRL2006_S}. 
In this state, 
while the magnetic moments in the triangular layers are polarized, 
the electric quadrupole moments in the kagom\'{e} layers behave as a 2D system. 
It can be regarded as a 2D frustrated system of quadrupole moments.}

{\color[rgb]{0,0,0} Here we also note that 
CMC simulations qualitatively reproduce 
the temperature dependence of the magnetic susceptibility along the $[111]$ direction~\cite{Kadowaki_MC_S}.
It shows a similar anomaly at $T_{\rm c}$ to that observed in experiments (Fig.~2(a) of the main text). 
This result supports the conclusion that Tb$_{2+x}$Ti$_{2-x}$O$_{7+y}$ can be basically represented by
the simple effective $S= 1/2$ Hamiltonian (Eq.(1) of the main text), although there could be non-negligible 
and complicated terms for a full description.}

\section{Treatment of the magnetic moment of the ground state doublet}
{\color[rgb]{0,0,0}In this section, we describe the treatment of the magnetic moment of the ground state doublet $\mu$
through the calculations 
and its effect to the estimation of parameters in Eq.~(1) of the main text.
It is known that} 
there are several reports on the estimation of $\mu$ for TTO:
$\mu = 5.1$~$\mu_{\rm B}$/Tb$^{3+}$ (Ref.~\cite{MirebeauPRB2007_S}), 
$4.8$~$\mu_{\rm B}$/Tb$^{3+}$ (Ref.~\cite{BertinPRB2012_S}),
$3.9$~$\mu_{\rm B}$/Tb$^{3+}$ (Ref.~\cite{ZhangPRB2014_S}), and
$5.4$~$\mu_{\rm B}$/Tb$^{3+}$ (Ref.~\cite{PrincepPRB2015_S}).
Therefore, there is some ambiguity in the choice of $\mu$.
This fact also affects the choice of $D_{\rm nn}$ for the calculations.
In this context, we adopted $\mu = 4.6$~$\mu_{\rm B}$/Tb$^{3+}$ in the calculations 
since it reproduces reasonably well the observed value of $M$ at the first kink of $\mu_{0}H_{1}$.
However, even when using other choices of $\mu$, 
the main conclusion of the present study does not alter. 
It only changes parameter values of $J_{\rm nn}$, $\delta$, and $q$ slightly:
for example, if we use $\mu = 5.3$~$\mu_{\rm B}$/Tb$^{3+}$ (i.e., $D_{\rm nn} = 0.63$~K),
we obtain $J_{\rm nn} = 0.9(1)$~K, and $(\delta, q) = (0.0\pm0.4, 1.0\pm0.3)$
for reproducing experimental results.
With these parameter values, we still obtain the PAF phase, although
the boundary between the SI and PAF phases should be changed slightly from 
that shown in Fig.~1(a) of the main text.

\section{Inelastic neutron scattering spectrum at temperatures above $T_{\rm c}$}
\begin{figure}[h]
\begin{center}
 \includegraphics[width=0.44\textwidth]{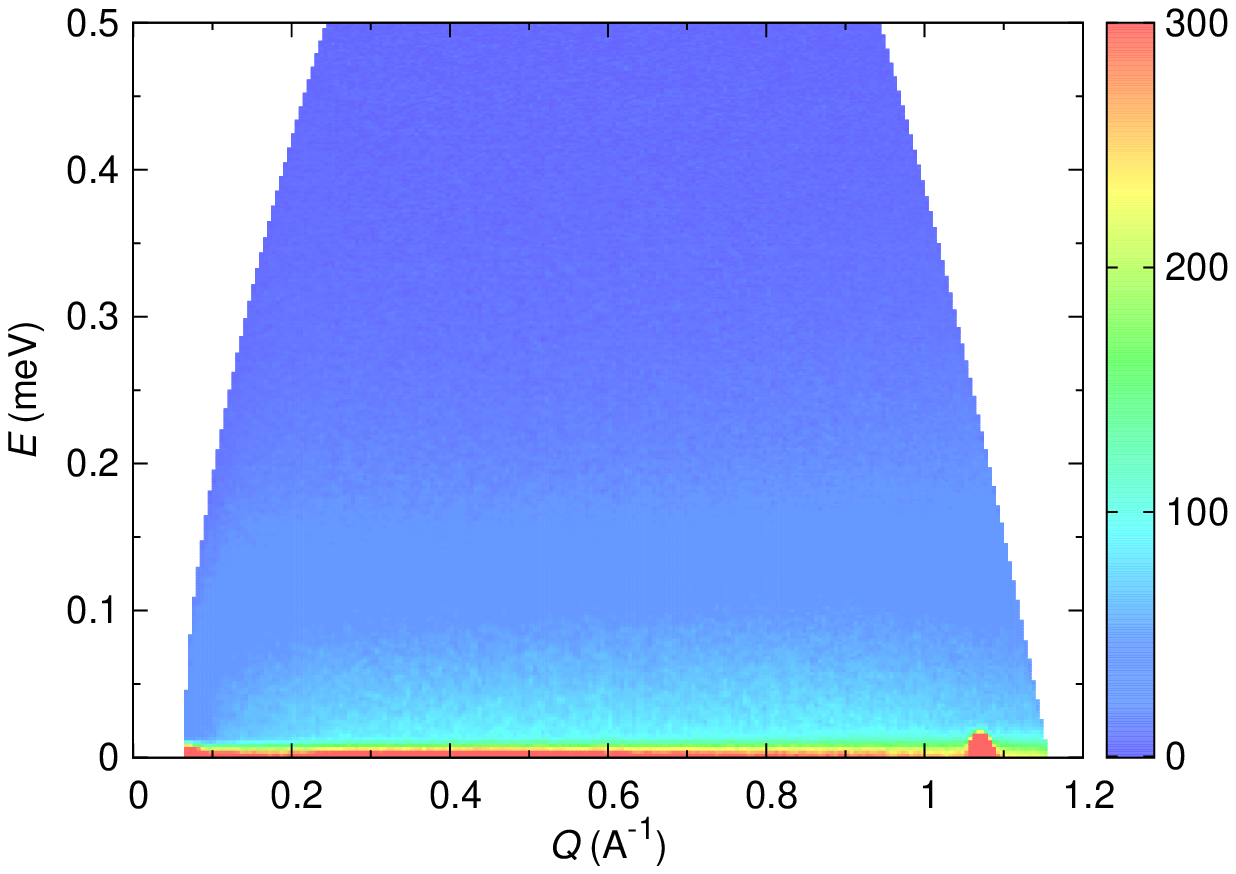}
\caption{
(Color online)
Neutron inelastic magnetic spectrum at $T = 0.7$~K (above $T_{\rm c} = 0.53$~K).
}
\label{fig.s2}
\end{center}
\end{figure}

{\color[rgb]{0,0,0}
In this section, we present the INS data of the Tb$_{2.005}$Ti$_{1.995}$O$_{7+y}$ powder sample 
above $T_{\rm c}$ and discuss a possibility of a local symmetry reduction which affects 
the splitting of the INS spectrum.

Figure~\ref{fig.s2} shows the INS spectrum at $T = 0.7$~K.
A quasielastic scattering spectrum is observed. 
This spectrum changes into an inelastic scattering spectrum with 
decreasing temperatures below $T_{\rm c}$ (see Fig.~3(a) of the main text and Ref.~\cite{TaniguchiPRB2013_S}).
This result indicates that the local symmetry reduction induced by the substitution of an element like $x$ of 
Tb$_{2+x}$Ti$_{2-x}$O$_{7+y}$ is a very minor effect or is absent 
for the case of Tb$_{2.005}$Ti$_{1.995}$O$_{7+y}$. This is because the local symmetry reduction often inevitably couples to 
the static Jahn-Teller effect which may lead to the splitting of the INS spectrum persisting to temperatures above $T_{\rm c}$.
Therefore, the result of Fig.~\ref{fig.s2} indicates 
that the splitting of the peak (Fig.~3(a) of the main text) is mainly induced by another effect.
We discussed it by considering the development of a pseudospin wave in the quadrupolar LRO state 
and demonstrated a similar spectrum by MF-RPA calculations 
(see the main text and Ref.~\cite{KadowakiSPIN2015_S} for the details).}

%

\end{document}